\documentclass{aip-cp}
\usepackage[numbers]{natbib}
\usepackage{rotating}
\usepackage[12pt]{moresize}
\usepackage{graphicx}

\begin{document}


\title{Emergent Magnetism at the 3$d$-5$d$ Interface: SrMnO$_3/$SrIrO$_3$}

\author{Sayantika Bhowal and Sashi Satpathy}

\affil {Department of Physics \& Astronomy, University of Missouri, Columbia, MO 65211, USA}
\maketitle

\begin{abstract}
Recent experiments have found new magnetic behaviors, which are different from the parent bulk materials, at the interfaces between 3$d$ and 5$d$ oxides such as SrMnO$_3$ (SMO) and SrIrO$_3$ (SIO). The system is of considerable interest due to the strong spin-orbit coupling in the 5$d$ materials on one hand and the double exchange physics in SMO on the other, which belongs to the class of the colossal magnetoresistive (CMR) manganites. In order to gain insight into the physics of the system, we have performed density-functional studies on a selected interface structure, viz., the  (SMO)$_1$(SIO)$_1$ superlattice, which has been experimentally grown and studied.
Our density-functional results show that the interfacial magnetism is controlled by a net charge transfer at the interface from the SIO to the SMO side, turning both of them into ferromagnetic metal from the original antiferromagnetic insulating state in the bulk.
The transferred electrons to the SMO side make it
ferromagnetic through the Anderson-Hasegawa double exchange interaction,
while the SIO part becomes ferromagnetic due to the doping of the half-filled Mott-Hubbard insulator as
suggested by the Nagaoka Theorem.
Our results are discussed  in the context of the experiments for the same structure.
\end{abstract}

\section{INTRODUCTION}
Transition metal oxides (TMO) are ideal playgrounds for the observation of the interplay between charge, spin, orbital and lattice degrees of freedom leading to many fascinating properties such as the metal-insulator transition \cite{Mott}, multiferroic properties \cite{multiferroic}, quantum spin liquid state \cite{QSL}, colossal magneto-resistance \cite{CMR}, unconventional superconductivity \cite{SC}, and the like.
While the physics of the 3$d$ TMO is governed by the strong Coulomb interaction, in the 5$d$ TMO, it is the large  spin-orbit coupling (SOC) that plays a dominating role. Thus the unique combination of strong Coulomb repulsion and SOC can
provide an important means of engineering the electronic and magnetic properties at the 3$d$-5$d$ interfaces,
with properties different from those in the bulk.
In fact, as we show in the current paper, a novel conducting and ferromagnetic (FM) region is predicted to occur at the interface, while both parent materials are antiferromagnetic (AFM) insulators.

A number of interface structures between the 3$d$ and 5$d$ TMO have been grown and studied recently \cite{Nichols,Matsuno,Wei}.
One of the notable example of these  is the (001) (SMO)$_m/$(SIO)$_n$ superlattice grown on the SrTiO$_3$ substrate \cite{Nichols}. 
In contrast to the parent oxides, where SMO is a G-type AFM insulator and SIO is a canted AFM insulator as found in $T = 0$ K density functional calculation \cite{Zeb}, magnetization and transport measurements of the interface reveal a FM ground state as well as a strong anomalous Hall effect \cite{Nichols}. 
Furthermore, X-ray absorption and X-ray magnetic circular dichroism spectroscopies \cite{Nichols} indicate a charge transfer from SIO to SMO, which has been interpreted to be due to the formation of molecular orbitals at the interface \cite{Okamoto}.

We show here the emergence of ferromagnetism at the interfacial layers of the 3$d$ SrMnO$_3$ (SMO) and 5$d$ SrIrO$_3$ (SIO) interface in agreement with the experimental observation \cite{Nichols} and unravel its mechanism.
Detail density-functional calculations are performed for the (001) (SMO)$_1$(SIO)$_1$ superlattice structure. 
We find the magnetism at the interfacial layers to be the outcome of the charge transfer from the SIO to the SMO part. 
This leads to  ferromagnetism in the electron doped SMO side, which is driven by the Anderson-Hasegawa double exchange (AHDE), while the ferromagnetism on the SIO side, which effectively becomes hole doped, is governed by the physics of the doped Mott-Hubbard insulator, where as suggested by the Nagaoka Theorem, a single hole in the half-filled Hubbard model turns the system into a FM metal from the AFM insulating state.

\section{STRUCTURE AND COMPUTATIONAL METHODS}

\begin{figure}[h]\label{fig1}
  \centerline{\includegraphics[width=200pt]{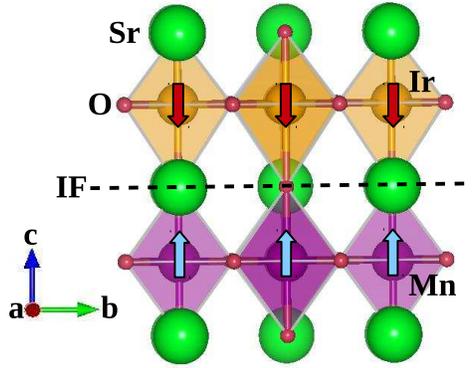}}
  \caption{ (color online) The ideal (001) (SMO)$_1$(SIO)$_1$ superlattice structure studied in the paper.
  Also indicated is the magnetic ground state obtained from our calculations, which is FM within the layers and AFM between the layers.}
\end{figure}

The bulk SMO crystallizes in the ideal cubic perovskite structure, 
while SIO crystallizes in the orthorhombic perovskite structure. 
In the present work, in order to construct the (SMO)$_1$(SIO)$_1$ superlattice, we have considered the idealized cubic structure ignoring the orthorhombic distortion of the iridate as shown in Fig. \ref{fig1}. 
The unit cell of the (SMO)$_1$(SIO)$_1$ superlattice contains two formula units of SMO and SIO. 
The in-plane lattice parameter ($a = b$) of the superlattice structure is fixed to the experimental lattice constant of the substrate SrTiO$_3$ (3.905 $\times \sqrt{2}$ \AA) while the out of plane lattice parameters are fixed to the respective experimental lattice constants of the bulk structures {\it i. e,} 3.80 \AA~for SMO \cite{Takeda} and 3.94 \AA~ for SIO \cite{Longo,Zhao}.

In the bulk, SMO is an AFM insulator. Experimentally, SIO is a paramagnetic metal which turns into a paramagnetic insulator below a transition temperature T$_{MI} \sim $ 44 K \cite{Zhao}. 
However, density-functional studies \cite{Zeb} suggest that at $T = 0$ it should be a canted AFM insulator, which is not inconsistent with the experimental measurements to the lowest temperatures. 
Our results will show that the two sides of the interface turn FM within the layers, while remaining AFM between the layers as indicated in Fig. \ref{fig1}.

Our first principles calculations based on density functional theory (DFT) have been performed using the plane-wave based projector augmented wave (PAW) \cite{PAW1,PAW2} method as implemented in the Vienna {\it ab initio} simulation package (VASP) \cite{vasp1,vasp2}. Exchange and correlation effects are treated within the generalized gradient approximation (GGA) \cite{gga} of Perdew-Burke-Ernzerhof including Hubbard $U$ \cite{U} and SOC. The Hubbard $U$ for Ir and Mn-$d$ states are taken to be $U$ = 2 eV and $U$ = 3 eV respectively following earlier authors \cite{Okamoto}. The kinetic energy cut-off of the plane wave basis was chosen to be 550 eV and a $\Gamma$ centered 6$\times$6$\times$4  k-mesh has been used for the Brillouin-zone integration.       

\section{RESULTS AND DISCUSSIONS}

 {\it DFT results for the bulk} --
 Before we discuss the results for the interface, we briefly indicate the results of our calculations for the two bulk materials.
 We find that the bulk cubic SMO has the $t_{2g}^3$ configuration and it is a G-type AFM insulator with magnetic moment 2.77 $\mu_B$/Mn, which is consistent with the experimental value of 2.6 $\pm$ 0.2 $\mu_B$ \cite{Takeda}. 
For SIO, the electronic structure is governed by the strong SOC of Ir,
 which splits the partially filled $t_{2g}$ states (filled with five electrons with the $d^5$ configuration)  into spin-orbit entangled completely filled J$_{\rm eff}$ = 3/2 quartet and half-filled J$_{\rm eff}$ = 1/2 doublet. 
 Orthorhombic SIO is a paramagnetic material with effective magnetic moment $\sim$ 0.12 $\mu_B$ \cite{Zhao}. 
Transport measurements show that SIO undergoes an insulator to metal transition at T$_{\rm MI}$ about 44 K with the low temperature structure being insulating \cite{Zhao}. Our DFT calculations ($T = 0$) show that SIO is a canted AFM  insulator in agreement with the previous report of first principles study on the orthorhombic SIO \cite{Zeb}. 
Indeed, experiments show that the paramagnetic metallic state is in close proximity to the AFM insulating state \cite{Zheng},
which is found to be the ground-state from the DFT calculations. 

{\it Magnetism at the interface} -- 
We have considered two magnetic configurations denoted as ``FM1" and ``AFM". 
In the ``FM1" configuration, the interaction within the layer is FM but it is AFM between the MnO$_2$ and IrO$_2$ layers (see Fig. \ref{fig1}) while for the ``AFM" configuration, both the intra-layer as well as the inter-layer interactions are AFM. 
The calculation of total energy within GGA+SOC+U, shows that ``FM1" is lower in energy (see Table \ref{tab1}) indicating the emergence of ferromagnetism at the interface consistent with the experimental report \cite{Nichols}.  

\begin{table} [h]\label{tab1}
\caption{ Total energies per (SMO)$_1$(SIO)$_1$ formula unit and charge transfer 
across the interface (electrons transferred from SIO to SMO side) per interface Ir atom. The ``FM1" structure is the ground state configuration as indicated in Fig. \ref{fig1}, 
while ``AFM" denotes the structure where the magnetic moments are antiferromagnetically aligned within the layers and also between the two layers. The total energy of the ``FM1" structure is set to zero.}
\centering
\tabcolsep60pt\begin{tabular}{c c c}
\hline
\hline
  Magnetic \ & $\Delta E$  & Charge    \\[0.3 ex]
  Structure \ &   (eV/fu)  & Transfer/Ir  \\
\hline
FM1   & 0 & 0.06 \\
AFM  &   0.12    & 0.04 \\
\hline\hline
\end{tabular}
\end{table}
     
\begin{figure}[t]\label{fig2}
\vspace{2 cm}
\centerline{\includegraphics[width=250pt]{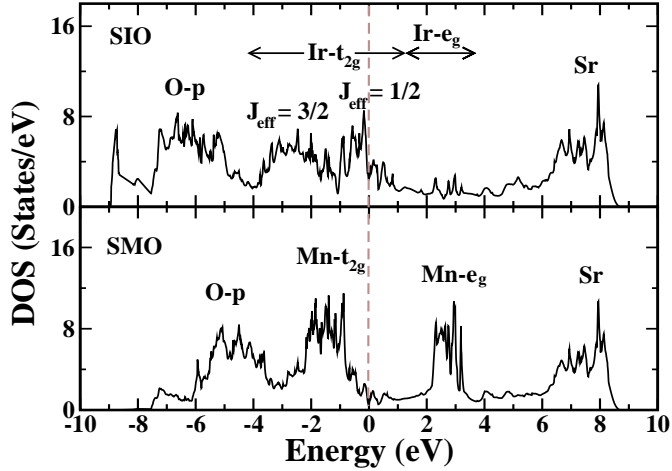}}
\caption{ (color online) The layer projected density of states (DOS) for the (SMO)$_1$(SIO)$_1$ superlattice with the magnetic ground-state structure as shown in Fig.\ref{fig1}. The figure indicates metallic behavior and a numerical integration of the DOS curves up to Fermi energy (taken  as the zero of energy) yields a transfer of electrons ($\sim$ 0.06 e$^-$) from the SIO layer to the MnO layer.}
\end{figure}

The layer projected density of states (DOS) corresponding to this ``FM1" configuration (see Fig. \ref{fig1}) is shown in Fig. \ref{fig2}. 
As we can see from Fig. \ref{fig2}, the interfacial SMO layer not only becomes FM but also metallic due to the transfer of electrons from the half-filled spin-orbit entangled J$_{\rm eff}$ = 1/2 state of the iridate to the empty Mn-$e_g^\uparrow$ states. 
The charge transfer across the interface may be computed by integrating the total DOS up to the Fermi energy ($E_F$) on the two sides, shown in Fig. \ref{fig2},
which however shows the projected charges within the atomic spheres and therefore must be appropriately renormalized. 
Approximately the same charge transfer is obtained if one integrates the partial DOS for the Ir and Mn atoms.
Considering the charge transfer from the SIO to the SMO side, we find that there is a transfer of 0.06 electrons per interface Ir atom for the ``FM1" structure. This leads to a charge transfer of $3.9 \times10^{13}$ e$^- / $  cm$^{2}$ across the interface, which is comparable to
the density of the 2DEG in the well-studied polar interface of LaAlO$_3$ and SrTiO$_3$ \cite{Ohtomo, Brinkman, Zoran}.
In the higher-energy AFM configuration, the computed charge transfer is a bit smaller as seen from Table \ref{tab1}.

\begin{figure}[t]\label{fig3}
  \centerline{\includegraphics[width=200pt]{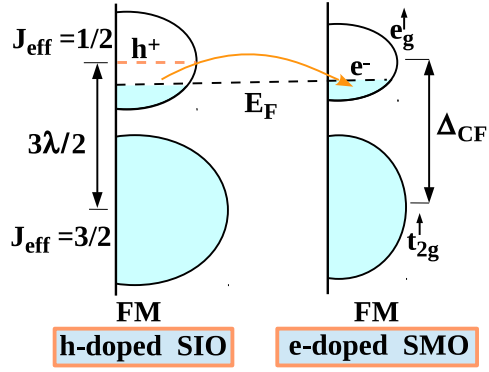}}
  \caption{ (color online) Schematic diagram showing the charge transfer at the interface from the half-filled J$_{\rm eff}$ = 1/2 state of SIO to the empty $e_g^\uparrow$ state of the SMO as obtained from the DFT results. The charge leakage from SIO makes it hole doped, while the SMO side becomes electron doped. 
  The doped carriers make both sides metallic as well as FM as discussed in the text.
}
\end{figure}

Regarding the magnetic moment, we find that on the SMO side, the  Mn spin moment is enhanced compared to the bulk value (3.23 $\mu_B$/Mn vs. 2.77 $\mu_B$/Mn in the bulk). 
As expected, the orbital moment at the Mn site is small, being about 0.03 $\mu_B$.
The electron leakage from SIO to SMO leaves holes on the SIO side and makes the SMO side electron-doped as indicated in Fig.\ref{fig3}. 
The hole-doped interfacial SIO also exhibits ferromagnetism and becomes metallic in contrast to the bulk properties. 
Both the spin and the orbital moment at the Ir site are found to be comparable, viz.,  0.21 and 0.16 $\mu_B$ respectively, which is expected for a strong spin-orbit coupled material. 

As already mentioned, ferromagnetism at the SMO/SIO interface is quite interesting in the sense that none of the parent oxides is FM in nature. In the following, we discuss a possible  mechanism for the ferromagnetism in the interface structure. 

\begin{figure}[t]\label{fig4}
\centerline{\includegraphics[width=170pt]{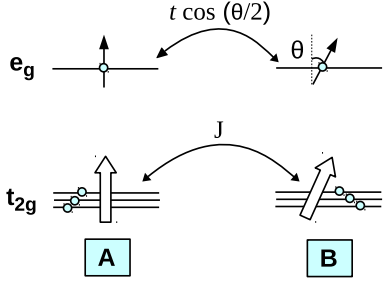}}
  \caption{ (color online) Illustration of the Anderson-Hasegawa double exchange interaction and the carrier-induced canted AFM state in a two-site system.
}
\end{figure}

{\it Anderson-Hasegawa double exchange and ferromagnetism in the SMO layer} --
The ferromagnetism at the SMO side can be explained in terms of the
double exchange developed for the colossal magneto-resistive (CMR) manganites. The leaked electrons into the manganite occupy the $e_g$ states and introduce ferromagnetism by AHDE \cite{Anderson}. In this mechanism (illustrated for a two-site system in Fig. \ref{fig4}), the superexchange of the core spins, fixed on the Mn sites ($t_{2g}$ spins), is overcome by the kinetic energy gain of the
doped carriers into the itinerant $e_g$ bands for sufficient amount of the carrier concentration $x$. For the case of infinite Hund's coupling $J_H$ (typically  $ \sim 1$ eV in the 3$d$ materials), the itinerant $e_g$ electrons can hop around the lattice if they are aligned in parallel with the core spins. Therefore hopping is allowed if the neighboring core spins are ferromagnetically aligned, while it is forbidden if the two spins are antiferromagnetically aligned. For intermediate alignment of the neighboring core spins with the ``canting" angle $\theta$, the effective hopping of the itinerant electron is given by the well-known Anderson-Hasegawa expression $t \cos (\theta/2)$ \cite{Anderson}. 

The canting angle is given by a competition between the AFM superexchange energy $J$  of the core spins (treated as classical spins) and the hopping energy of the doped electrons, which is described by the double-exchange Hamiltonian
\begin{equation}\label{AH}
{\cal H} = - t  \cos (\theta/2)   \sum_{ \langle ij \rangle }   \  (c_{i\uparrow}^{\dagger} c_{j\uparrow} + H. c.) 
+ \sum_{ \langle ij \rangle} J  \ \hat S_i \cdot \hat S_j,  
\end{equation}
where we have considered a two sublattice structure with the core spins of the two sublattices are FM within the sublattice and are aligned with the canting angle $\theta$ between the two sublattices, so that for the FM structure $\theta = 0$,
while for the AFM structure $\theta = \pi$, $c_{i\uparrow}^{\dagger} $ creates an electron at the site $i$ with spin aligned to the core spin $\vec S_i$ at that site, and the sum is over all distinct set of   nearest neighbors.
The Hund's energy being infinity, the corresponding spin state $c_{i\downarrow}^{\dagger} $ has infinite energy and does not play a role in this simple picture.

As pointed out by de Gennes in a seminal paper \cite{DeGennes}, the double exchange Hamiltonian Equation (\ref{AH}) turns the original AFM state to a canted AFM state and eventually to a FM state with increasing carrier concentration $x$. For small $x$, the electrons occupy the band bottom $E_b = -2z |t| \cos (\theta/2)$, where the number of nearest neighbor $z = 4$ corresponding to the present case of the Mn square lattice in the SMO layer. The canting angle $\theta$ is obtained by minimizing the total energy
\begin{equation}
E = E_b x + z J \cos \theta
\end{equation}
with respect to $\theta$, which immediately yields the result
\begin{equation}  \label{angle}
\theta = 2 \cos^{-1}   \big( \frac{ |t| x} {2 J}  \big) . 
\end{equation}

In the absence of charge leakage ($x = 0$), super-exchange leads to the AFM state with $\theta = \pi$. Now, with the increase in charge leakage ($x$) into the SMO, the strength of the double exchange interaction increases and eventually at a certain critical doping concentration ($x_c$) double exchange dominates over the existing super-exchange interaction. At this point, the total energy stabilizes at a canting angle $\theta \neq \pi$ leading to a canted AFM state. Further increase in $x$, leads to a FM state with $\theta = 0$. 

Clearly, as indicated by the expression (\ref{angle}),  we recover the AFM state for $x = 0$, while the system turns FM beyond the critical value given by
$x_c  = 2 J / |t| \approx 0.13$, where $J \approx 10$ meV and $|t| \approx 0.15$ eV for SMO. This result was obtained in the limit $J_H \rightarrow \infty$, but as shown in our earlier work \cite{Nanda} the critical concentration diminishes by about a factor of two if a finite $J_H \approx 1$ eV is used, resulting in the value $x_c \approx 0.07$. That is about $7 \%$ doped charge carrier would turn the material FM. From our DFT calculations, we found a charge transfer of about $6\%$ across the interface, which  though not more than the critical value predicted from the simplified model, is quite close to it. This suggests the AHDE as the mechanism that turns the SMO part FM, caused by the charge transfer across the interface.

{\it SIO layer as a hole doped Mott-Hubbard insulator} --
The ferromagnetism in the SIO part can be described in terms of the hole-doped Mott-Hubbard insulator. The charge transfer from SIO to SMO leads to hole doping. As we have mentioned earlier, at  T = 0,  SIO is a canted AFM insulator. Now, according to the Nagaoka theorem \cite{Nagaoka}, in the strong-coupling Hubbard model ($U / W \rightarrow \infty$), a single hole destroys the AFM insulating state of the undoped half-filled Hubbard model, turning it into a FM metal. 
For a smaller $U$, a finite  but non-zero concentration of holes is needed to turn the system into FM.
Exact results do not exist in this case, however mean-field calculations of the 
 ground state phase diagram of the doped Hubbard model show that the FM state is favored when the hole  concentration exceeds a critical value \cite{Jamshid} (typically a few percent depending on the magnitude of $U/W$). This explains the observed ferromagnetism in the SIO layer. 

{\it Prediction for the (SMO)$_n$(SIO)$_n$ superlattices} --
Recently, in addition to the (SMO)$_1$(SIO)$_1$ superlattice, experimenters have systematically grown and studied thicker superlattices of the type (SMO)$_n$(SIO)$_n$ with $n > 1$. 
Based on the insight gained from our studies, we can predict the magnetism at the interface for these superlattices. 
With the increase in $n$, the electrons are not only leaked into the first interfacial SMO layer but also into the following layers away from the interface, leading to a reduced electron doping concentration in SMO. In fact, for the $ n \geq 2$  structure,
the charge leakage is expected to be nearly half of the $n = 1$ structure studied in the paper, since in the latter, charge in the SMO layer is transferred from two of the adjacent layers in the interface.
The hole concentration on the SIO side is similarly reduced as the SMO layer is situated only at one side as opposed to the (SMO)$_1$(SIO)$_1$ structure, where the SMO layers are at both sides of SIO. 
The reduced charge transfer is not then expected to exceed the critical value $x_c$,
so that the  doped carriers are unable to induce sufficient double exchange in order to overcome the existing AFM interaction.
Similarly, the smaller hole concentration on the SIO side may not exceed the critical value to produce   ferromagnetism.
The final result may be a canted AFM state driven by double exchange producing a reduced, but non-zero net FM moment in the 
superlattice structure.
Indeed, recent magnetic measurements \cite{Nichols} show that ferromagnetism is gradually reduced in going from $ n = 1$ to 3, and it completely disappears for $n \geq 4$, consistent with our argument.            

\section{SUMMARY}

In summary, we have performed density-functional studies on the recently grown SIO/SMO interfaces in order to gain insight into the behavior of interfaces between 3$d$ and 5$d$ materials. Concrete calculations were performed for the (SMO)$_1$(SIO)$_1$ superlattice grown along the (001) direction. The calculations show that there is a significant amount charge leakage of $3.9 \times10^{13}$ e$^- / $  cm$^{2}$ (0.06 e$^-$ per interface Ir) across the interface from the SIO to the SMO side, making the former hole doped and the latter electron doped. 

The transferred charge plays an important role in altering the magnetic interactions near the interface. 
The doped electrons turn the SMO part metallic and ferromagnetic via the well-known AHDE mechanism. The hole doped SIO part, on the other hand, behaves like a hole doped Hubbard system and becomes FM also. Both mechanisms require a critical amount of doped carriers ($x_c \sim$ a few percent, depending on the Hamiltonian parameters such as the $U/W$ ratio, Hund's energy $J_H$, etc.) to turn the system FM, and the transferred charge in the (SMO)$_1$(SIO)$_1$ structure exceeds this value. 

Extending this to the (SMO)$_n$(SIO)$_n$ superlattices with $n > 1$, we argued that the ferromagnetism at the interface gradually becomes weaker with increasing layer thickness $n$, and after a critical value of $n \sim 4$, it completely disappears, which qualitatively agrees with recent experiments \cite{Nichols}.  The reason for this is that the leaked carriers penetrate deeper into the bulk, making it below the critical concentration $x_c$ for any layer, so that the doped carriers are not sufficient in number to alter the original antiferromagnetism. 
The effect allows for the engineering of the interface magnetism by tuning the amount of charge transfer across the interface, which can presumably be done by external means such as gate voltage and strain. It would be valuable if such experiments can be performed in the future in order to assess their effects on the magnetism at the interface.

\section{ACKNOWLEDGMENT}    
We thank U.S. Department of Energy, Office of Basic Energy Sciences, Division of Materials Sciences and Engineering under Grant No. DE-FG02-00ER45818 for financial support.

\nocite{*}
\bibliographystyle{aipnum-cp}%
\bibliography{AIP}%

\begin{thebibliography}{28}%
\makeatletter
\providecommand \@ifxundefined [1]{%
 \@ifx{#1\undefined}
}%
\providecommand \@ifnum [1]{%
 \ifnum #1\expandafter \@firstoftwo
 \else \expandafter \@secondoftwo
 \fi
}%
\providecommand \@ifx [1]{%
 \ifx #1\expandafter \@firstoftwo
 \else \expandafter \@secondoftwo
 \fi
}%
\providecommand \natexlab [1]{#1}%
\providecommand \enquote  [1]{``#1''}%
\providecommand \bibnamefont  [1]{#1}%
\providecommand \bibfnamefont [1]{#1}%
\providecommand \citenamefont [1]{#1}%
\providecommand \href@noop [0]{\@secondoftwo}%
\providecommand \href [0]{\begingroup \@sanitize@url \@href}%
\providecommand \@href[1]{\@@startlink{#1}\@@href}%
\providecommand \@@href[1]{\endgroup#1\@@endlink}%
\providecommand \@sanitize@url [0]{\catcode `\$12\catcode `\&12\catcode
  `\#12\catcode `\^12\catcode `\_12\catcode `\%12\relax}%
\providecommand \@@startlink[1]{}%
\providecommand \@@endlink[0]{}%
\providecommand \url  [0]{\begingroup\@sanitize@url \@url }%
\providecommand \@url [1]{\endgroup\@href {#1}{\urlprefix }}%
\providecommand \urlprefix  [0]{URL }%
\providecommand \Eprint [0]{\href }%
\providecommand \doibase [0]{http://dx.doi.org/}%
\providecommand \selectlanguage [0]{\@gobble}%
\providecommand \bibinfo  [0]{\@secondoftwo}%
\providecommand \bibfield  [0]{\@secondoftwo}%
\providecommand \translation [1]{[#1]}%
\providecommand \BibitemOpen [0]{}%
\providecommand \bibitemStop [0]{}%
\providecommand \bibitemNoStop [0]{.\EOS\space}%
\providecommand \EOS [0]{\spacefactor3000\relax}%
\providecommand \BibitemShut  [1]{\csname bibitem#1\endcsname}%
\let\auto@bib@innerbib\@empty
\bibitem [{\citenamefont {Imada}, \citenamefont {Fujimori},\ and\ \citenamefont
  {Tokura}(1998)}]{Mott}%
  \BibitemOpen
  \bibfield  {author} {\bibinfo {author} {\bibfnamefont {M.}~\bibnamefont
  {Imada}}, \bibinfo {author} {\bibfnamefont {A.}~\bibnamefont {Fujimori}}, \
  and\ \bibinfo {author} {\bibfnamefont {Y.}~\bibnamefont {Tokura}},\
  }\href@noop {} {\bibfield  {journal} {\bibinfo  {journal} {Rev. Mod. Phys.}\
  }\textbf {\bibinfo {volume} {70}},\ \ \bibinfo {pages} {1039} (\bibinfo
  {year} {1998})}\BibitemShut {NoStop}%
\bibitem [{\citenamefont {Eerenstein}, \citenamefont {Mathur},\ and\
  \citenamefont {Scott}(2006)}]{multiferroic}%
  \BibitemOpen
  \bibfield  {author} {\bibinfo {author} {\bibfnamefont {W.}~\bibnamefont
  {Eerenstein}}, \bibinfo {author} {\bibfnamefont {N.~D.}\ \bibnamefont
  {Mathur}}, \ and\ \bibinfo {author} {\bibfnamefont {J.~F.}\ \bibnamefont
  {Scott}},\ }\href@noop {} {\bibfield  {journal} {\bibinfo  {journal}
  {Nature}\ }\textbf {\bibinfo {volume} {442}},\ \ \bibinfo {pages} {759}
  (\bibinfo {year} {2006})}\BibitemShut {NoStop}%
\bibitem [{\citenamefont {Lawler}\ \emph {et~al.}(2008)\citenamefont {Lawler},
  \citenamefont {Paramekanti}, \citenamefont {Kim},\ and\ \citenamefont
  {Balents}}]{QSL}%
  \BibitemOpen
  \bibfield  {author} {\bibinfo {author} {\bibfnamefont {M.~J.}\ \bibnamefont
  {Lawler}}, \bibinfo {author} {\bibfnamefont {A.}~\bibnamefont {Paramekanti}},
  \bibinfo {author} {\bibfnamefont {Y.~B.}\ \bibnamefont {Kim}}, \ and\
  \bibinfo {author} {\bibfnamefont {L.}~\bibnamefont {Balents}},\ }\href@noop
  {} {\bibfield  {journal} {\bibinfo  {journal} {Phys. Rev. Lett.}\ }\textbf
  {\bibinfo {volume} {101}},\ \ \bibinfo {pages} {197202} (\bibinfo {year}
  {2008})}\BibitemShut {NoStop}%
\bibitem [{\citenamefont {Salamon}\ and\ \citenamefont {Jaime}(2001)}]{CMR}%
  \BibitemOpen
  \bibfield  {author} {\bibinfo {author} {\bibfnamefont {M.~B.}\ \bibnamefont
  {Salamon}}\ and\ \bibinfo {author} {\bibfnamefont {M.}~\bibnamefont
  {Jaime}},\ }\href@noop {} {\bibfield  {journal} {\bibinfo  {journal} {Rev.
  Mod. Phys.}\ }\textbf {\bibinfo {volume} {73}},\ \ \bibinfo {pages} {583}
  (\bibinfo {year} {2001})}\BibitemShut {NoStop}%
\bibitem [{\citenamefont {Norman}(2011)}]{SC}%
  \BibitemOpen
  \bibfield  {author} {\bibinfo {author} {\bibfnamefont {M.~R.}\ \bibnamefont
  {Norman}},\ }\href@noop {} {\bibfield  {journal} {\bibinfo  {journal}
  {Science}\ }\textbf {\bibinfo {volume} {332}},\ \ \bibinfo {pages} {196}
  (\bibinfo {year} {2011})}\BibitemShut {NoStop}%
\bibitem [{\citenamefont {Nichols}\ \emph {et~al.}(2016)\citenamefont
  {Nichols}, \citenamefont {Gao}, \citenamefont {Lee}, \citenamefont {Meyer},
  \citenamefont {Freeland}, \citenamefont {Lauter}, \citenamefont {Yi},
  \citenamefont {Liu}, \citenamefont {Haskel}, \citenamefont {Petrie},
  \citenamefont {Guo}, \citenamefont {Herklotz}, \citenamefont {Lee},
  \citenamefont {Z.Ward}, \citenamefont {Eres}, \citenamefont {Fitzsimmons},\
  and\ \citenamefont {Lee}}]{Nichols}%
  \BibitemOpen
  \bibfield  {author} {\bibinfo {author} {\bibfnamefont {J.}~\bibnamefont
  {Nichols}}, \bibinfo {author} {\bibfnamefont {X.}~\bibnamefont {Gao}},
  \bibinfo {author} {\bibfnamefont {S.}~\bibnamefont {Lee}}, \bibinfo {author}
  {\bibfnamefont {T.~L.}\ \bibnamefont {Meyer}}, \bibinfo {author}
  {\bibfnamefont {J.~W.}\ \bibnamefont {Freeland}}, \bibinfo {author}
  {\bibfnamefont {V.}~\bibnamefont {Lauter}}, \bibinfo {author} {\bibfnamefont
  {D.}~\bibnamefont {Yi}}, \bibinfo {author} {\bibfnamefont {J.}~\bibnamefont
  {Liu}}, \bibinfo {author} {\bibfnamefont {D.}~\bibnamefont {Haskel}},
  \bibinfo {author} {\bibfnamefont {J.~R.}\ \bibnamefont {Petrie}}, \bibinfo
  {author} {\bibfnamefont {E.-J.}\ \bibnamefont {Guo}}, \bibinfo {author}
  {\bibfnamefont {A.}~\bibnamefont {Herklotz}}, \bibinfo {author}
  {\bibfnamefont {D.}~\bibnamefont {Lee}}, \bibinfo {author} {\bibfnamefont
  {T.}~\bibnamefont {Z.Ward}}, \bibinfo {author} {\bibfnamefont
  {G.}~\bibnamefont {Eres}}, \bibinfo {author} {\bibfnamefont {M.~R.}\
  \bibnamefont {Fitzsimmons}}, \ and\ \bibinfo {author} {\bibfnamefont {H.~N.}\
  \bibnamefont {Lee}},\ }\href@noop {} {\bibfield  {journal} {\bibinfo
  {journal} {Nat. Commun.}\ }\textbf {\bibinfo {volume} {7}},\ \ \bibinfo
  {pages} {12721} (\bibinfo {year} {2016})}\BibitemShut {NoStop}%
\bibitem [{\citenamefont {Matsuno}\ \emph {et~al.}(2015)\citenamefont
  {Matsuno}, \citenamefont {Ihara}, \citenamefont {Yamamura}, \citenamefont
  {Wadati}, \citenamefont {Ishii}, \citenamefont {Shankar}, \citenamefont
  {Kee},\ and\ \citenamefont {Takag}}]{Matsuno}%
  \BibitemOpen
  \bibfield  {author} {\bibinfo {author} {\bibfnamefont {J.}~\bibnamefont
  {Matsuno}}, \bibinfo {author} {\bibfnamefont {K.}~\bibnamefont {Ihara}},
  \bibinfo {author} {\bibfnamefont {S.}~\bibnamefont {Yamamura}}, \bibinfo
  {author} {\bibfnamefont {H.}~\bibnamefont {Wadati}}, \bibinfo {author}
  {\bibfnamefont {K.}~\bibnamefont {Ishii}}, \bibinfo {author} {\bibfnamefont
  {V.~V.}\ \bibnamefont {Shankar}}, \bibinfo {author} {\bibfnamefont {H.-Y.}\
  \bibnamefont {Kee}}, \ and\ \bibinfo {author} {\bibfnamefont
  {H.}~\bibnamefont {Takag}},\ }\href@noop {} {\bibfield  {journal} {\bibinfo
  {journal} {Phys. Rev. Lett.}\ }\textbf {\bibinfo {volume} {114}},\ \ \bibinfo
  {pages} {247209} (\bibinfo {year} {2015})}\BibitemShut {NoStop}%
\bibitem [{\citenamefont {Wei}\ and\ \citenamefont {Seiji}(2015)}]{Wei}%
  \BibitemOpen
  \bibfield  {author} {\bibinfo {author} {\bibfnamefont {F.}~\bibnamefont
  {Wei}}\ and\ \bibinfo {author} {\bibfnamefont {Y.}~\bibnamefont {Seiji}},\
  }\href@noop {} {\bibfield  {journal} {\bibinfo  {journal} {J. Phys. Conf.
  Ser.}\ }\textbf {\bibinfo {volume} {592}},\ \unskip\ \bibinfo {pages}
  {012139--012145} (\bibinfo {year} {2015})}\BibitemShut {NoStop}%
\bibitem [{\citenamefont {Zeb}\ and\ \citenamefont {Kee}(2012)}]{Zeb}%
  \BibitemOpen
  \bibfield  {author} {\bibinfo {author} {\bibfnamefont {M.~A.}\ \bibnamefont
  {Zeb}}\ and\ \bibinfo {author} {\bibfnamefont {H.-Y.}\ \bibnamefont {Kee}},\
  }\href@noop {} {\bibfield  {journal} {\bibinfo  {journal} {Phys. Rev. B}\
  }\textbf {\bibinfo {volume} {86}},\ \ \bibinfo {pages} {085149} (\bibinfo
  {year} {2012})}\BibitemShut {NoStop}%
\bibitem [{\citenamefont {Okamoto}\ \emph {et~al.}(2017)\citenamefont
  {Okamoto}, \citenamefont {Nichols}, \citenamefont {Sohn}, \citenamefont
  {Kim}, \citenamefont {Noh},\ and\ \citenamefont {Lee}}]{Okamoto}%
  \BibitemOpen
  \bibfield  {author} {\bibinfo {author} {\bibfnamefont {S.}~\bibnamefont
  {Okamoto}}, \bibinfo {author} {\bibfnamefont {J.}~\bibnamefont {Nichols}},
  \bibinfo {author} {\bibfnamefont {C.}~\bibnamefont {Sohn}}, \bibinfo {author}
  {\bibfnamefont {S.~Y.}\ \bibnamefont {Kim}}, \bibinfo {author} {\bibfnamefont
  {T.~W.}\ \bibnamefont {Noh}}, \ and\ \bibinfo {author} {\bibfnamefont
  {H.~N.}\ \bibnamefont {Lee}},\ }\href@noop {} {\bibfield  {journal} {\bibinfo
   {journal} {Nano Lett.}\ }\textbf {\bibinfo {volume} {17}},\ \unskip\
  \bibinfo {pages} {2126--2130} (\bibinfo {year} {2017})}\BibitemShut {NoStop}%
\bibitem [{\citenamefont {Takeda}\ and\ \citenamefont {Ohara}(1974)}]{Takeda}%
  \BibitemOpen
  \bibfield  {author} {\bibinfo {author} {\bibfnamefont {T.}~\bibnamefont
  {Takeda}}\ and\ \bibinfo {author} {\bibfnamefont {S.}~\bibnamefont {Ohara}},\
  }\href@noop {} {\bibfield  {journal} {\bibinfo  {journal} {J. Phys. Soc.
  Jpn.}\ }\textbf {\bibinfo {volume} {37}},\ \ \bibinfo {pages} {275} (\bibinfo
  {year} {1974})}\BibitemShut {NoStop}%
\bibitem [{\citenamefont {Longo}, \citenamefont {Kafalas},\ and\ \citenamefont
  {Arnott}(1971)}]{Longo}%
  \BibitemOpen
  \bibfield  {author} {\bibinfo {author} {\bibfnamefont {J.~M.}\ \bibnamefont
  {Longo}}, \bibinfo {author} {\bibfnamefont {J.~A.}\ \bibnamefont {Kafalas}},
  \ and\ \bibinfo {author} {\bibfnamefont {R.~J.}\ \bibnamefont {Arnott}},\
  }\href@noop {} {\bibfield  {journal} {\bibinfo  {journal} {J. Solid State
  Chem.}\ }\textbf {\bibinfo {volume} {3}},\ \ \bibinfo {pages} {174} (\bibinfo
  {year} {1971})}\BibitemShut {NoStop}%
\bibitem [{\citenamefont {Zhao}\ \emph {et~al.}(2008)\citenamefont {Zhao},
  \citenamefont {Yang}, \citenamefont {Yu}, \citenamefont {Li}, \citenamefont
  {Yu}, \citenamefont {Fang}, \citenamefont {Chen},\ and\ \citenamefont
  {Jin}}]{Zhao}%
  \BibitemOpen
  \bibfield  {author} {\bibinfo {author} {\bibfnamefont {J.~G.}\ \bibnamefont
  {Zhao}}, \bibinfo {author} {\bibfnamefont {L.~X.}\ \bibnamefont {Yang}},
  \bibinfo {author} {\bibfnamefont {Y.}~\bibnamefont {Yu}}, \bibinfo {author}
  {\bibfnamefont {F.~Y.}\ \bibnamefont {Li}}, \bibinfo {author} {\bibfnamefont
  {R.~C.}\ \bibnamefont {Yu}}, \bibinfo {author} {\bibfnamefont
  {Z.}~\bibnamefont {Fang}}, \bibinfo {author} {\bibfnamefont {L.~C.}\
  \bibnamefont {Chen}}, \ and\ \bibinfo {author} {\bibfnamefont {C.~Q.}\
  \bibnamefont {Jin}},\ }\href@noop {} {\bibfield  {journal} {\bibinfo
  {journal} {J. Appl. Phys.}\ }\textbf {\bibinfo {volume} {103}},\ \ \bibinfo
  {pages} {103706.} (\bibinfo {year} {2008})}\BibitemShut {NoStop}%
\bibitem [{\citenamefont {Blochl}(1994)}]{PAW1}%
  \BibitemOpen
  \bibfield  {author} {\bibinfo {author} {\bibfnamefont {P.~E.}\ \bibnamefont
  {Blochl}},\ }\href@noop {} {\bibfield  {journal} {\bibinfo  {journal} {Phys.
  Rev. B}\ }\textbf {\bibinfo {volume} {50}},\ \ \bibinfo {pages} {17953}
  (\bibinfo {year} {1994})}\BibitemShut {NoStop}%
\bibitem [{\citenamefont {Kresse}\ and\ \citenamefont {Joubert}(1999)}]{PAW2}%
  \BibitemOpen
  \bibfield  {author} {\bibinfo {author} {\bibfnamefont {G.}~\bibnamefont
  {Kresse}}\ and\ \bibinfo {author} {\bibfnamefont {D.}~\bibnamefont
  {Joubert}},\ }\href@noop {} {\bibfield  {journal} {\bibinfo  {journal} {Phys.
  Rev. B}\ }\textbf {\bibinfo {volume} {59}},\ \ \bibinfo {pages} {1758}
  (\bibinfo {year} {1999})}\BibitemShut {NoStop}%
\bibitem [{\citenamefont {Kresse}\ and\ \citenamefont {Hafner}(1993)}]{vasp1}%
  \BibitemOpen
  \bibfield  {author} {\bibinfo {author} {\bibfnamefont {G.}~\bibnamefont
  {Kresse}}\ and\ \bibinfo {author} {\bibfnamefont {J.}~\bibnamefont
  {Hafner}},\ }\href@noop {} {\bibfield  {journal} {\bibinfo  {journal} {Phys.
  Rev. B}\ }\textbf {\bibinfo {volume} {47}},\ \ \bibinfo {pages} {558}
  (\bibinfo {year} {1993})}\BibitemShut {NoStop}%
\bibitem [{\citenamefont {Kresse}\ and\ \citenamefont
  {Furthm{\"u}ller}(1996)}]{vasp2}%
  \BibitemOpen
  \bibfield  {author} {\bibinfo {author} {\bibfnamefont {G.}~\bibnamefont
  {Kresse}}\ and\ \bibinfo {author} {\bibfnamefont {J.}~\bibnamefont
  {Furthm{\"u}ller}},\ }\href@noop {} {\bibfield  {journal} {\bibinfo
  {journal} {Phys. Rev. B}\ }\textbf {\bibinfo {volume} {54}},\ \ \bibinfo
  {pages} {11169} (\bibinfo {year} {1996})}\BibitemShut {NoStop}%
\bibitem [{\citenamefont {Perdew}, \citenamefont {Burke},\ and\ \citenamefont
  {Ernzerhof}(1996)}]{gga}%
  \BibitemOpen
  \bibfield  {author} {\bibinfo {author} {\bibfnamefont {J.~P.}\ \bibnamefont
  {Perdew}}, \bibinfo {author} {\bibfnamefont {K.}~\bibnamefont {Burke}}, \
  and\ \bibinfo {author} {\bibfnamefont {M.}~\bibnamefont {Ernzerhof}},\
  }\href@noop {} {\bibfield  {journal} {\bibinfo  {journal} {Phys. Rev. Lett.}\
  }\textbf {\bibinfo {volume} {77}},\ \ \bibinfo {pages} {3865} (\bibinfo
  {year} {1996})}\BibitemShut {NoStop}%
\bibitem [{\citenamefont {Anisimov}, \citenamefont {Zaanen},\ and\
  \citenamefont {Andersen}(1991)}]{U}%
  \BibitemOpen
  \bibfield  {author} {\bibinfo {author} {\bibfnamefont {V.~I.}\ \bibnamefont
  {Anisimov}}, \bibinfo {author} {\bibfnamefont {J.}~\bibnamefont {Zaanen}}, \
  and\ \bibinfo {author} {\bibfnamefont {O.~K.}\ \bibnamefont {Andersen}},\
  }\href@noop {} {\bibfield  {journal} {\bibinfo  {journal} {Phys. Rev. B}\
  }\textbf {\bibinfo {volume} {44}},\ \ \bibinfo {pages} {943} (\bibinfo {year}
  {1991})}\BibitemShut {NoStop}%
\bibitem [{\citenamefont {Zheng}\ \emph {et~al.}(2016)\citenamefont {Zheng},
  \citenamefont {Terzic}, \citenamefont {Ye}, \citenamefont {Wan},
  \citenamefont {Wang}, \citenamefont {Wang}, \citenamefont {Wang},
  \citenamefont {Schlottmann}, \citenamefont {Yuan},\ and\ \citenamefont
  {Cao}}]{Zheng}%
  \BibitemOpen
  \bibfield  {author} {\bibinfo {author} {\bibfnamefont {H.}~\bibnamefont
  {Zheng}}, \bibinfo {author} {\bibfnamefont {J.}~\bibnamefont {Terzic}},
  \bibinfo {author} {\bibfnamefont {F.}~\bibnamefont {Ye}}, \bibinfo {author}
  {\bibfnamefont {X.~G.}\ \bibnamefont {Wan}}, \bibinfo {author} {\bibfnamefont
  {D.}~\bibnamefont {Wang}}, \bibinfo {author} {\bibfnamefont {J.}~\bibnamefont
  {Wang}}, \bibinfo {author} {\bibfnamefont {X.}~\bibnamefont {Wang}}, \bibinfo
  {author} {\bibfnamefont {P.}~\bibnamefont {Schlottmann}}, \bibinfo {author}
  {\bibfnamefont {S.~J.}\ \bibnamefont {Yuan}}, \ and\ \bibinfo {author}
  {\bibfnamefont {G.}~\bibnamefont {Cao}},\ }\href@noop {} {\bibfield
  {journal} {\bibinfo  {journal} {Phys. Rev. B}\ }\textbf {\bibinfo {volume}
  {93}},\ \ \bibinfo {pages} {235157} (\bibinfo {year} {2016})}\BibitemShut
  {NoStop}%
\bibitem [{\citenamefont {Ohtomo}\ and\ \citenamefont {Hwang}(2004)}]{Ohtomo}%
  \BibitemOpen
  \bibfield  {author} {\bibinfo {author} {\bibfnamefont {A.}~\bibnamefont
  {Ohtomo}}\ and\ \bibinfo {author} {\bibfnamefont {H.}~\bibnamefont {Hwang}},\
  }\href@noop {} {\bibfield  {journal} {\bibinfo  {journal} {Nature (London)}\
  }\textbf {\bibinfo {volume} {427}},\ \ \bibinfo {pages} {423} (\bibinfo
  {year} {2004})}\BibitemShut {NoStop}%
\bibitem [{\citenamefont {Brinkman}\ \emph {et~al.}(2007)\citenamefont
  {Brinkman}, \citenamefont {Huijben}, \citenamefont {van Zalk}, \citenamefont
  {Huijben}, \citenamefont {Zeitler}, \citenamefont {Maan}, \citenamefont
  {van~der Wiel}, \citenamefont {Rijnders}, \citenamefont {Blank},\ and\
  \citenamefont {Hilgenkamp}}]{Brinkman}%
  \BibitemOpen
  \bibfield  {author} {\bibinfo {author} {\bibfnamefont {A.}~\bibnamefont
  {Brinkman}}, \bibinfo {author} {\bibfnamefont {M.}~\bibnamefont {Huijben}},
  \bibinfo {author} {\bibfnamefont {M.}~\bibnamefont {van Zalk}}, \bibinfo
  {author} {\bibfnamefont {J.}~\bibnamefont {Huijben}}, \bibinfo {author}
  {\bibfnamefont {U.}~\bibnamefont {Zeitler}}, \bibinfo {author} {\bibfnamefont
  {J.~C.}\ \bibnamefont {Maan}}, \bibinfo {author} {\bibfnamefont {W.~G.}\
  \bibnamefont {van~der Wiel}}, \bibinfo {author} {\bibfnamefont
  {G.}~\bibnamefont {Rijnders}}, \bibinfo {author} {\bibfnamefont {D.~H.~A.}\
  \bibnamefont {Blank}}, \ and\ \bibinfo {author} {\bibfnamefont
  {H.}~\bibnamefont {Hilgenkamp}},\ }\href@noop {} {\bibfield  {journal}
  {\bibinfo  {journal} {Nature Mater.}\ }\textbf {\bibinfo {volume} {6}},\ \
  \bibinfo {pages} {493} (\bibinfo {year} {2007})}\BibitemShut {NoStop}%
\bibitem [{\citenamefont {Popovi\'c}, \citenamefont {Satpathy},\ and\
  \citenamefont {Martin}(2008)}]{Zoran}%
  \BibitemOpen
  \bibfield  {author} {\bibinfo {author} {\bibfnamefont {Z.~S.}\ \bibnamefont
  {Popovi\'c}}, \bibinfo {author} {\bibfnamefont {S.}~\bibnamefont {Satpathy}},
  \ and\ \bibinfo {author} {\bibfnamefont {R.~M.}\ \bibnamefont {Martin}},\
  }\href@noop {} {\bibfield  {journal} {\bibinfo  {journal} {Phys. Rev. Lett.}\
  }\textbf {\bibinfo {volume} {101}},\ \ \bibinfo {pages} {256801} (\bibinfo
  {year} {2008})}\BibitemShut {NoStop}%
\bibitem [{\citenamefont {Anderson}\ and\ \citenamefont
  {Hasegawa}(1955)}]{Anderson}%
  \BibitemOpen
  \bibfield  {author} {\bibinfo {author} {\bibfnamefont {P.}~\bibnamefont
  {Anderson}}\ and\ \bibinfo {author} {\bibfnamefont {H.}~\bibnamefont
  {Hasegawa}},\ }\href@noop {} {\bibfield  {journal} {\bibinfo  {journal}
  {Phys. Rev.}\ }\textbf {\bibinfo {volume} {100}},\ \ \bibinfo {pages} {675}
  (\bibinfo {year} {1955})}\BibitemShut {NoStop}%
\bibitem [{\citenamefont {de~Gennes}(1960)}]{DeGennes}%
  \BibitemOpen
  \bibfield  {author} {\bibinfo {author} {\bibfnamefont {P.-G.}\ \bibnamefont
  {de~Gennes}},\ }\href@noop {} {\bibfield  {journal} {\bibinfo  {journal}
  {Phys. Rev.}\ }\textbf {\bibinfo {volume} {118}},\ \ \bibinfo {pages} {141}
  (\bibinfo {year} {1960})}\BibitemShut {NoStop}%
\bibitem [{\citenamefont {Nanda}, \citenamefont {Satpathy},\ and\ \citenamefont
  {Springborg}(2007)}]{Nanda}%
  \BibitemOpen
  \bibfield  {author} {\bibinfo {author} {\bibfnamefont {B.~R.~K.}\
  \bibnamefont {Nanda}}, \bibinfo {author} {\bibfnamefont {S.}~\bibnamefont
  {Satpathy}}, \ and\ \bibinfo {author} {\bibfnamefont {M.~S.}\ \bibnamefont
  {Springborg}},\ }\href@noop {} {\bibfield  {journal} {\bibinfo  {journal}
  {Phys. Rev. Lett.}\ }\textbf {\bibinfo {volume} {98}},\ \ \bibinfo {pages}
  {216804} (\bibinfo {year} {2007})}\BibitemShut {NoStop}%
\bibitem [{\citenamefont {Nagaoka}(1966)}]{Nagaoka}%
  \BibitemOpen
  \bibfield  {author} {\bibinfo {author} {\bibfnamefont {Y.}~\bibnamefont
  {Nagaoka}},\ }\href@noop {} {\bibfield  {journal} {\bibinfo  {journal} {Phys.
  Rev.}\ }\textbf {\bibinfo {volume} {147}},\ \ \bibinfo {pages} {392}
  (\bibinfo {year} {1966})}\BibitemShut {NoStop}%
\bibitem [{\citenamefont {Kurdestany}\ and\ \citenamefont
  {Satpathy}(2017)}]{Jamshid}%
  \BibitemOpen
  \bibfield  {author} {\bibinfo {author} {\bibfnamefont {J.~M.}\ \bibnamefont
  {Kurdestany}}\ and\ \bibinfo {author} {\bibfnamefont {S.}~\bibnamefont
  {Satpathy}},\ }\href@noop {} {\bibfield  {journal} {\bibinfo  {journal}
  {Phys. Rev. B}\ }\textbf {\bibinfo {volume} {96}},\ \ \bibinfo {pages}
  {085132} (\bibinfo {year} {2017})}\BibitemShut {NoStop}%
\end{thebibliography}%

\end{document}